\documentclass[
amssymb,nobibnotes,aps,pre,showpacs]{revtex4}

\usepackage{hyperref}
\usepackage{graphicx}
\usepackage{color}
\usepackage{bm}

\usepackage{graphicx,amssymb,amstext,amsmath}

\usepackage{epsfig}

\usepackage{epstopdf}

\usepackage{color,graphics}

\newcommand\be{\begin{equation}}
\newcommand\e{\end{equation}}
\newcommand\ba{\begin{eqnarray}}
\newcommand\ay{\end{eqnarray}}
\newcommand\nn{\nonumber}

\newcommand\cab{\text{C}_{2_1}^0}
\newcommand\cac{\text{C}_{2_2}^0}
\newcommand\cad{\text{C}_{2_3}^0}

\newcommand\ccb{\text{C}_{2_1}^2}
\newcommand\ccc{\text{C}_{2_2}^2}
\newcommand\ccd{\text{C}_{2_3}^2}
\newcommand\pd{\partial}
\newcommand\eps{\epsilon}

\newcommand\ntz{n_{i2}^{(0)}}

\newcommand{\vtz}{v_{i2}^{(0)}}
\newcommand{\ptz}{\phi_2^{(0)}}

\newcommand\mtz{n_{e2}^{(0)}}

\newcommand\pss{|\psi|^2}

\newcommand\vfrac{\frac{ak}{\omega}}
\newcommand\utz{v_{e2}^{(0)}}
\newcommand\cam{\text{C}_{2_4}^0}
\newcommand\cau{\text{C}_{2_5}^0}
\newcommand\ccm{\text{C}_{2_4}^2}
\newcommand\ccu{\text{C}_{2_5}^2}

\begin{document}

\title{Ion-Acoustic Envelope Modes in a Degenerate Relativistic Electron-Ion Plasma}

\author{M. McKerr$^1$, F. Haas$^2$,  I. Kourakis$^1$}

\affiliation{ $^1$ Centre for Plasma Physics, School of Mathematics and Physics, Queen's University Belfast, BT7 1NN Belfast, Northern Ireland, UK\\
$^2$ Instituto de F\'{\i}sica, Universidade Federal do Rio Grande do Sul,  Av. Bento Gon\c{c}alves 9500, Porto Alegre, RS, Brazil}

\date{\today}

\begin{abstract}
A self-consistent relativistic two-fluid model is proposed for one-dimensional electron-ion plasma dynamics.  A multiple scales perturbation technique is employed, leading to an evolution equation for the wave envelope, in the form of a nonlinear Schr\"odinger type equation (NLSE). The inclusion of relativistic effects is shown to introduce density-dependent factors, not present in the non-relativistic case - in the conditions for modulational instability. The role of relativistic effects on the linear dispersion laws and on envelope soliton solutions of the NLSE is discussed.
\end{abstract}


\maketitle

\section{Introduction}

The elucidation of the dynamics of ultra-high density plasmas  in one-dimensional (1D) geometry \cite{Giamarchi} is recognized as a challenging area of research among researchers in the last decade.
Dense plasmas occur in ``extreme'' astrophysical environments, such as white dwarfs or neutron stars \cite{Canuto1, Canuto2, Carbonaro} and in the  core of giant planets (e.g., Jovian planets) \cite{Koester1990, Shapiro1983, Chabrier2002, Fortov2009}.  Such plasmas may also occur in the next generation of laser-based matter compression schemes \cite{Fortov2009,Shukla2011}.
The topic has gained momentum recently, thanks to its relevance to high-power laser-assisted energy production (fusion) research, and in particular to the target normal sheath acceleration (TNSA) mechanism \cite{Passoni} during the irradiation of solid targets with a high-intensity laser beam
\cite{Thiele}. Other applications of (1D) degenerate plasmas include: the dense quantum diode \cite{Shukla}, the electron-hole plasma in quantum wires \cite{Barak}, the 1D fermionic Luttinger liquid \cite{Imambekov}, and 1D semiconductor quantum wells \cite{Haas1}, to mention a few.

In such extreme plasma environments,  magnetic fields can be extremely strong, effectively varying over many orders of magnitude,  from a few kilogauss to gigagauss (or even
petagauss) in white dwarfs (neutron stars, respectively), hence effectively confining particle motion to one dimension (1D). On the other hand, temperatures can be quite high, comparable to fusion plasma  ($\sim10^{8}$
K) \cite{Shukla2011}. In such conditions, quantum degeneracy and
relativity effects are ubiquitous, since the de-Broglie wavelength may approach, or even exceed,  the inter-particle (fermion) distance.
At
extremely high densities, the electron Fermi energy $E_{Fe0}$ can exceed by far than thermal energy, hence the electron thermal pressure may be
negligible, compared to the Fermi degeneracy pressure; the latter arises due to the combined effect of Pauli's exclusion principle and Heisenberg's uncertainty principle.

From a nonlinear dynamical point of view, ultrahigh-density plasmas pose a real challenge; their rich and varied dynamics may sustain a wide range of excitations, from breather-mode oscillations in 1D semiconductors \cite{Haas1} and Lagrangian structures in dense 1D plasmas \cite{Ghosh} to 1D nonlinear envelope modes in dense electron-positron-ion plasmas \cite{McKerr}, quasi-1D solitons \cite{Tercas} and wakefields in quantum wires \cite{Ali1}, among others.
It may be added that the study of the dynamics of 1D plasmas is certainly not restricted to dense systems only. In Ref. \onlinecite{Melrose}, kinetic theoretical arguments have been employed to found the possibility of reconnection between Langmuir and Alfv\'en modes in a strongly-magnetized, non-degenerate, relativistic pair plasma.

For ultra-high plasma densities, relativistic effects need to be included in plasma modeling, since the relativistic parameter $p_{F}/mc$ \cite{Salpeter} ($p_F$ is the Fermi momentum,  $m$ is the electron mass and $c$ is the light speed) acquires large values, thus modifying the equation of state and hence the dynamical plasma profile.
Many authors have considered the problem of re\-lativistically dense plasma before, from different angles.
Such problems as the formation of elec\-tro\-sta\-tic shocks within an electron-ion plasma \cite{Eliasson}, the exis\-tence of arbitrary-amplitude solitary structures \cite{Akbari} and small-amplitude envelope modes \cite{Rahman} within an electron-positron-ion plasma have  been studied in the past. Stationary profile electrostatic pulses and Langmuir-type excitations have been investigated in Refs. \onlinecite{MMKPRE2014}  and \onlinecite{HaasPPCF}, respectively.
 However, the majority of works tacitly apply Chandrasekhar's (three-dimensional, 3D) equation of state \cite{Chandrasekhar}, a reasonable assumption since the environments under consideration in the above (e.g. white dwarf stars) certainly occupy three dimensions. What we aim for, in the study at hand, is an understanding of envelope modes in a dense, 1D plasma, such as is used as a model for the study of target normal sheath acceleration (TNSA) \cite{Passoni}.

We shall here focus on a relativistic one-dimensional model for dense plasmas. Our aim is to propose a self-consistent fully relativistic theoretical framework for low (ionic) frequency electrostatic modulated envelope structures propagating in unmagnetized electron-ion plasma. The model comprises an inertialess electron fluid, which is described by a quantum-mechanical degenerate distribution function, and a classical inertial ion fluid. A fully relativistic fluid model is adopted for both components.
The use of Fermi-Dirac statistics in the description of the electron fluid forces us to countenance the exclusion principle. In the case of high densities, a significant overlap of the electrons' position-wavefunctions leads to a pressure, which, according to Pauli's exclusion principle, exists to resist degeneracy as would occur if two electrons were to share the same state. Such a pressure results in a considerable (density-dependent) momentum near the Fermi surface (the surface in momentum-space below which all states are occupied) of the electron gas, the magnitude of which demands a relativistic treatment. To this end, an equation of state is employed which is similar to that of Chandrasekhar \cite{Chandrasekhar}, but is essentially that of the one-dimensional ``water-bag" distribution
\cite{Katsouleas}. Unlike the original  Chandrasekhar equation of state \cite{Chandrasekhar}, which was developed for one-dimensional (1D) propagation (in fact, in the radial direction) within a spherical-symmetric geometry, our equations of state is suitable for for modeling strictly 1D propagation dynamics \cite{MMKPRE2014}.

The electrons will be treated as ``cold", so as to avail of the zero-temperature Fermi-Dirac distribution. Such an approximation is justified under certain conditions which depend on the density, entering the algebraic description via the relativistic electron Fermi energy $E_{Fe, rel}$, viz.
\be k_BT_e\ll E_{Fe, rel} = m_e c^2 \sqrt{1+p_{Fe}^2/m_e^2c^2}-m_ec^2. \label{relativisticassumption}\e
Here, $k_B$ is Boltzmann's constant, $T_e$ is the electron (thermal) temperature, $c$ is the speed of light \textit{in vacuo} and $p_{Fe}$ and $m_e$ are respectively the local Fermi momentum and the rest mass of the electron. The local Fermi momentum is expressed in terms of the local density, $n_e$, and Planck's constant, $h$, as $p_{Fe}=hn_e/4$.

The layout of this article goes as follows. In the next Section II, a self-consistent, relativistic fluid model is introduced. The evolution equation for the plasma state variables are then scaled, and a dimensionless system in presented in Section III. A multiple scale perturbation technique is employed in Section IV, and  then analyzed in the lowest (linear) and higher (nonlinear) order(s) in Sections V and VI, respectively. The modulational wavepacket profile is outlined in Section VII. Localized envelope structures are introduced in Section VIII. A parametric analysis is presented in Section IX, and the results are summarized in the concluding Section X.

\section{Fluid model}

We are interested in investigating ion dynamics in a degenerate relativistic plasma. We shall adopt the quantum hydrodynamic description \cite{Haasbook}, by introducing a fluid model which is described in the following. The ion fluid is described by its particle (number) density, $n_i$, and velocity, $v_i$. It is a ``cold", fully-ionized fluid of singly-charged, positive ions, whose dynamics is dominated by electric forces deriving from an electrostatic potential, $\phi(x,t)$. A magnetic field has not been considered, for the sake of simplicity.

The electron fluid constitutes an inertialess background to the ion dynamics. It is characterized by a number density $n_e$ and a fluid velocity, $v_e$, directed along the $x-$axis.  The electrons are considered to be relativistically degenerate and therefore the appropriate equation of state to govern their motion is provided by the expression for relativistic degeneracy pressure in one dimension \cite{Chavanis, Katsouleas}:
\be P_e=\frac{2m_e^2c^3}{h}\left[\xi_e(\xi_e^2+1)^{1/2}-\sinh^{-1} \xi_e \, ,\right] \label{Pequation} \e
where $\xi_e=p_{Fe}/(m_ec)$ is a dimensionless parameter measuring the effect of relativistic electron effects.
The latter equation of state is a consequence of the Pauli exclusion principle and is valid for arbitrary strength of relativistic effects. Note that an expansion of the pressure (\ref{Pequation}) for low density- $\xi_0\ll1$- yields the non-relativistic 1D Fermi pressure, $P_e=2E_{Fe0}n_0(n_e/n_0)^3/3$. Similarly, an ultrarelativistic ($\xi_e\gg1$)  approximation is found to be  $P_e=cp_{Fe0}n_0(n_e/n_0)^2/2$, where $p_{Fe0}=hn_0/4$.

The model comprises five equations, namely the fluid-dynamical equations expressing continuity (number density conservation) and momentum conservation for the ion and electron fluid(s), with the system closed by Poisson's equation for the electrostatic potential $\phi$, which essentially couples the dynamical variables to one another.
\ba&&\frac{\pd (\gamma_in_i)}{\pd t}+\frac{\pd}{\pd x}(\gamma_i n_i v_i)=0 \, , \nn\\
&&\frac{\pd (\gamma_e n_e)}{\pd t} + \frac{\pd}{\pd x}(\gamma_e n_e v_e)=0 \, ,  \nn\\
&&\frac{\pd (\gamma_i v_i)}{\pd t} + v_i\frac{\pd (\gamma_i v_i)}{\pd x}+\frac{e}{m_i}\frac{\pd\phi}{\pd x}=0 \, ,  \nn\\
&&e\frac{\pd\phi}{\pd x}-\frac{\gamma_e}{n_e}\left(\frac{\pd P_e}{\pd x}+\frac{v_e}{c^2}\frac{\pd P_e}{\pd t}\right)=0\nn\\
&&\frac{\pd^2\phi}{\pd x^2}+\frac{e}{\eps_0}(\gamma_in_i-\gamma_en_e)=0 \, \label{relfluidmodel} .\ay
Note that electron inertia has be neglected, according to the underlying assumptions of our model, as discussed above.
Adopting the electrostatic approximation, we have suppressed (neglected) magnetic field generation, hence the remaining Maxwell relations were omitted.
As expected in a relativistic model, the factor $\gamma_{e,i}=1/\sqrt{1-v_{e,i}^2/c^2}$ appears in the fluid-dynamical equations, as a result of Lorentz transformations and relations between quantities, such as the electron and ion number density (functions), between different inertial frames.

It is understood that the validity of our model equations (\ref{relfluidmodel}) above, assumes that assumption (\ref{relativisticassumption}) holds, i.e. for sufficiently high density.

\section{Dimensionless Model}

It is appropriate to derive a dimensionless model, by scaling by appropriate quantities.
A natural speed scale
in our physical problem is the characteristic quantity
\( c_s=\left({2 E_{Fe0}}/{m_i}\right)^{1/2} \, \)
where
\( E_{Fe0}={h^2n_0^2}/{(32 m_e)}\)
is the non-relativistic electron Fermi energy: this  is the equivalent of the ion ``sound speed" in classical plasma dynamics. Accordingly, wel adopt the following scaling:
\begin{eqnarray}
\nonumber
x &\rightarrow& \frac{\omega_{pi}x}{c_s} \,, \quad t \rightarrow \omega_{pi}t \,,\\
n_{e,i} &\rightarrow& \frac{n_{e,i}}{n_0} \,, \quad v_{e,i} \rightarrow \frac{v_{e,i}}{c_s} \,, \quad \phi \rightarrow \frac{e\phi}{m_i c_{s}^2} \,.
\end{eqnarray}
Note that $n_{e0}=n_{i0} = n_0$ from the quasi-neutrality condition (obtained upon  considering Poisson's relation at equilibrium).
Finally, a natural pressure scale $P_0 = e \phi_0 n_0$ is considered.
The evolution equations take the form:
\ba&&\frac{\pd \gamma_in_i}{\pd t}+\frac{\pd}{\pd x}(\gamma_in_iv_i)=0 \,,\nn\\
&&\frac{\pd \gamma_en_e}{\pd t}+\frac{\pd}{\pd x}(\gamma_en_ev_e)=0 \,, \nn\\
&&\frac{\pd \gamma_iv_i}{\pd t}+v_i\frac{\pd \gamma_i v_i}{\pd x}+\frac{\pd\phi}{\pd x}=0 \,, \nn\\
&&\frac{\pd\phi}{\pd x} - \frac{\gamma_e n_{e}}{\sqrt{1 + \xi_{0}^2 n_{e}^2}}\left(\frac{\pd n_e}{\pd x} + \alpha v_e \frac{\pd n_e}{\pd t}\right) = 0 \,, \nn\\
&&\frac{\pd^2\phi}{\pd x^2}+\gamma_in_i-\gamma_en_e=0 \,, \ay
where
\be
\gamma_{e,i} = \left(1 - \frac{c_{s}^2}{c^2}\,v_{e,i}^2\right)^{-1/2} \,.
\e
With this scaling, there is only one free parameter left: $\xi_0=hn_0/4m_ec$. The electron Fermi energy can be expressed as $E_{Fe0}=m_ec^2\xi_0^2/2$ and so can $\alpha$
as a result:
\be \alpha = \frac{c_s^2}{c^2}=\frac{2E_{Fe0}}{m_ic^2}=\frac{m_e\xi_0^2}{m_i} \, .
\e
We may, where appropriate, still retain  the notation for $\alpha$
below, recalling (rather than substituting with) the exact expression above, for the sake of  analytical tractability.

Concluding this Section, we note that the essential physics of our model is elegantly ``hidden" in the parameter $\xi_0$, which incorporates the relativistic effect, here manifested in terms of the (high) plasma density.

\section{Multiscale perturbation scheme}

A multiple-scales technique will be employed in the following \cite{Kourakis}. We anticipate a solution which comprises a fast carrier wave and a slowly-evolving envelope amplitude:
\be u\sim u(X_1,X_2,...,T_1,T_2)e^{i(kX_0-\omega T_0)}\nn\e
where $T_r=\eps^r t$ and $X_r=\eps^r x$; $\eps>0$ is a small, free parameter (it is independent of $X_r$ and $T_r$).

The state functions are expanded around their equilibrium values as
\ba n_i&\approx& 1+\eps n_{i1}+\eps^2n_{i2}+\eps^3 n_{i3}\nn\\
n_e&\approx& 1+\eps n_{e1}+\eps^2n_{e2}+\eps^3n_{e3}\nn\\
v_i&\approx& \eps v_{1}+\eps^2v_{2}+\eps^3 v_{3}\nn\\
v_e&\approx& \eps v_{e1}+\eps^2 v_{e2}+\eps^3 v_{e3}\nn\\
\phi&\approx& \eps \phi_1+\eps^2\phi_2+\eps^3\phi_3\ay

Furthermore, each of the functions is decomposed into Fourier components; for instance, for the velocity contribution in order $\eps^n$:
\be u_n=\sum_{l=-n}^nu_n^{(l)}e^{i l (k X_0 - \omega T_0)} \, . \e
This relation holds $\forall n = 1, 2, 3, ...$, hence
\begin{itemize}

\item $l = -1, 0, 1$ \ \ \ \ \ \ \ \ \ \ \ for $n=1$,

\item  $l = -2, -1, 0, 1, 2$ \ \ for $n=2$,

\end{itemize}

and so on.
Since these functions are real-valued, it must be imposed that
\be u_n^{(-r)}=\bar u_n^{(r)} \, . \nn\e

Upon substituting into the model equations (3) above, and then isolating successive contributions (orders in $\eps$), this perturbation/expansion scheme yields a system of polynomials in $\eps$ whose coefficients are required to vanish independently, since $\eps$ is free (arbitrary-valued). For any given value of $n$ ($=1, 2, ...$), these coefficients can be decomposed into their separate harmonics, expressed by the second index $l$ (taking values from $-n$ to $n$).  Each decomposition suggests a relation to be imposed between its constituent variables, which provides the solution for the given harmonic (amplitude). These expressions for the harmonics are then fed into the next order in $n$, and so on an so forth. The tedious, but straightforward algebraic procedure is presented in detail in Ref. \onlinecite{Kourakis}.

As an example, consider Poisson's equation at the second order of $\eps$:
\be\frac{\pd^2\phi_2}{\pd X_0^2}+2\frac{\pd^2\phi_1}{\pd X_0\pd X_1}+n_{i2}-n_{e2}+\frac{\alpha }{2}(v_{i1}^2-v_{e1}^2)=0 \, . \nn\e
This can be split into equations for the ``zeroth", first and second harmonics respectively:
\ba && \ntz-\mtz+\alpha\left(v_{i1}^{(1)}v_{i1}^{(-1)}-v_{e1}^{(1)}v_{e1}^{(-1)}\right)=0\nn\\
&&-k^2\phi_2^{(1)}+2ik\frac{\pd\phi_1^{(1)}}{\pd X_1}+\left(n_{i2}^{(1)}-n_{e2}^{(1)}\right)=0\nn\\
&&-4k^2\phi_2^{(2)}+\left(n_{i2}^{(2)}-n_{e2}^{(2)}\right)+\frac{\alpha }{2}\left({v_{i1}^{(1)}}^2-{v_{e1}^{(1)}}^2\right)=0 \, . \nn\ay
Analogous equations are obtained at all expansion and harmonic order(s), thus providing explicit solutions for the harmonic amplitudes. The tedious details of the algebraic procedure are omitted here: in the following, we shall provide the main steps. The relevant expressions for the harmonic amplitudes are reported in the Appendix.

\section{Linear response and dispersion relation}

At first order, there is only a first harmonic to investigate. The equations are presented below.
\ba &&-\omega n_{i1}+kv_{i1}=0 \, , \nn\\
&&-\omega n_{e1}+k v_{e1}=0 \, , \nn\\
&&-\omega v_{i1}+k\phi_1=0 \, , \nn\\
&&-\sqrt{1+\xi_0^2}\phi_1+n_{e1}=0 \, , \nn\\
&&-k^2\phi_1+n_{i1}-n_{e1}=0 \, . \ay
The electrons' equation of motion is used to eliminate $n_{e1}$ from Poisson's relation. The electrons' equation of continuity is used to find $v_{e1}$, but contains no other state variables.

The remaining three equations (the equations of continuity and of motion for the ions and Poisson's equation) can be arranged as follows:
 \be\left(\begin{array}{ccc}-\omega&k&0\\0&-\omega&k\\b&0&-(c_1+k^2)\end{array}\right)\left(\begin{array}{c} n_{i1}\\v_{i1}\\ \phi_1\end{array}\right)=\vec 0 \, , \nn\e
where
\be c_1=\sqrt{1+\xi_0^2} \, . \e
The vanishing determinant condition for a non-trivial solution to exist leads to the dispersion relation:
\be\omega^2=\frac{k^2}{c_1+k^2}\label{disp} \, .\e
Note that $c_1$ (defined above) essentially regulates the long-wavelength behavior of electrostatic waves, since $\omega \simeq k/\sqrt{c_1}$ for $k \ll 1$, while $\omega \simeq 1$ for $k \gg 1$.

\begin{figure}[h!]

\includegraphics[width=2.9in]{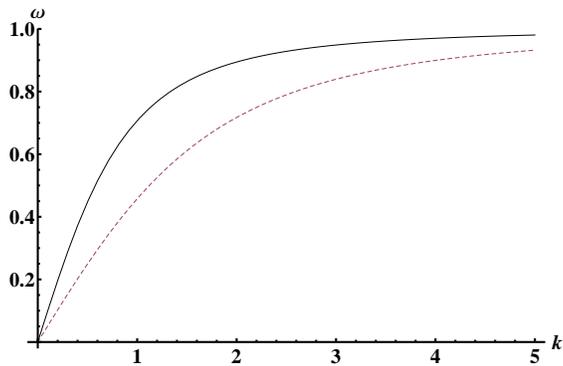}\\
\caption{(Color online) Plot of the frequency $\omega(k)$ for
density values: $n_0 = 10^{11} m^{-1}$ (continuous upper line) and $n_0 = 3 \times 10^{12} m^{-1}$ (dashed lower line).}
\label{figure1}
\end{figure}

Two important quantities are to be retained (and distinguished) at this stage, of high importance in the dynamics of modulated wavepackets: the phase speed ($v_{ph}=\omega/k$) and the group velocity ($v_g=d\omega/dk$).  These can directly be obtained as functions of the wavenumber $k$: see Fig.\ref{figure1}. Caution must be exercised in the interpretation of the plots, as the scales are density-dependent, so curves with different equilibrium density are not presented on the same scale. This will be discussed further below.

The solutions for the first harmonic amplitudes, as obtained in this order, can be expressed as functions of the electrostatic potential (leading-order disturbance) amplitude $\phi_1^{(1)} = \psi$ as:
\ba
n_{i1}^{(1)}&=&
\frac{k}{\omega} v_{i1}^{(1)} = \frac{k^2}{\omega^2}\psi \, , \quad \,
v_{e1}^{(1)} = \frac{\omega}{k} n_{e1}^{(1)}  = \frac{c_1\omega}{k}\psi \, . \quad  \ay

\section{Nonlinear treatment}

We may now consider the evolution equations at the next (second) order in the expansion parameter $\epsilon$.
The five equations for the first harmonic at the second order of $\eps$ are reduced to three by use of the expressions derived for the first order. The three remaining equations can again be expressed as a matrix equation, this time the vector formed of $n_{i2}^{(1)}$, $v_{i2}^{(1)}$ and $\phi_2^{(1)}$ being acted upon by the same matrix operator as before. The degeneracy of this matrix forces the following condition on the first-order amplitudes:
\be \frac{\pd\psi}{\pd T_1}+v_g\frac{\pd\psi}{\pd X_1}=0 \, .\e
This condition essentially ensures that secular terms (which would potentially lead to divergent solutions) are eliminated.
It results that $\psi$ depends on the first-order variables as $\psi = \psi(X_1 - v_g T_1)$, suggesting that the envelope moves at the group velocity, $v_g=d\omega/dk$.

We can freely set $\phi_2^{(1)}$ to zero. Using the linear relations between these second-order quantities, we find:
\ba\phi_2^{(1)}&=&0\nn\\
n_{i2}^{(1)}&=&-2ik\frac{\pd\psi}{\pd X_1}\nn\\
v_{i2}^{(1)}&=&-i\omega\frac{\pd\psi}{\pd X_1}\nn\\
n_{e2}^{(1)}&=&0\nn\\
v_{e2}^{(1)}&=&\frac{ic_1}{\omega }\left(1-\frac{\omega}{k}v_g\right)\frac{\pd\psi}{\pd X_1} \, . \ay
The second-harmonic and zeroth-harmonic amplitudes are found to be proportional to $\psi^2$ and $\pss$ respectively. The exact formulae are given in the Appendix.

Applying the same method to the five equations in the first harmonic at \emph{third} order in $\eps$, we obtain a consistency condition in the form of a \emph{non-linear Schr\"odinger Equation} (NLSE)
\be i\frac{\pd\psi}{\pd \tau}+P\frac{\pd^2\psi}{\pd \xi^2}+Q\pss\psi=0 . \label{NLS}\e
Here, the time and space variables are $\tau = T_2 = \eps^2 t$ and $\xi = X_1 - v_f T_1 = \eps (x - v_g t)$, respectively. The coefficient $P=  d^2\omega/2 dk^2$ gives rise to dispersion, where  the coefficient $Q$ represents cubic nonlinearity:  the full expression for $Q$ is given in the Appendix, owing to its length.

It may be appropriate to discuss the long-wavelength (small $k$) behavior  of the coefficients $P$ and $Q$, by deriving approximations to their respective expressions; these are:
\ba P&\approx& -\frac{3k}{2c_1^{3/2}},\nn\\
Q&\approx&\frac{1}{12c_1^{3/2}k}\left(4+3\xi_0^2-2\alpha c_1\right)\left(4+3\xi_0^2-\alpha c_1\right).\label{pqsmall}\ay
Note that $P < 0$ in this region, while $Q$ can be shown to be positive, thus ensuring stability for large wavelengths, as will be discussed below.

\section{Modulational stability analysis}

Let us adopt a periodic reference solution, $\psi_0=a_0e^{iQa_0^2t}$. To derive a dispersion relation for a periodic disturbance to this solution, we append small, real corrections of the same magnitude to both the amplitude and the phase:
\be \psi_0\mapsto (a_0+a_1)e^{i(Qa_0^2 t+b_1)}.\e

Since the corrections take real values, the NLSE (\ref{NLS}) can be separated into real and imaginary parts after inserting the perturbed solution. The two equations thus obtained are
\ba &&-\frac{\pd b_1}{\pd t}+P\frac{\pd^2a_1}{\pd x^2}+2Qa_0^2=0\nn\\
&&\frac{\pd a_1}{\pd t}+P\frac{\pd^2b_1}{\pd x^2}=0 .\ay

The corrections are periodic, taking the form $a_1 =Ae^{i(\kappa x-\Omega t)}+c.c.$ and $b=Be^{i(\kappa x-\Omega t)}+c.c.$, where $A$ and $B$ are complex and $\Omega$ and $\kappa$ are real. Inserting this into the equations above yields a pair of simultaneous equations in $A$ and $B$, the consistency condition of which is the required dispersion relation:
\be \Omega^2=\left(P\kappa\right)^2\left(1-\frac{2Qa_0^2}{P\kappa^2}\right). \label{growth}\e
It is evident that $\Omega$ is always real if $Q/P>0$. If $Q/P<0$, then $\Omega$ will be imaginary below a certain threshold,
\be \kappa< \kappa_{crit}=a_0\sqrt\frac{2Q}{P}, \label{defkappacritical} \e
up until which point the solution will be unstable. This interval is dependent on the value, $k$, of the wavenumber of the solution. The growth rate attains its maxinum at $\kappa_{max}=a_0\sqrt{Q/P}$. This mechanism is equivalent to the Benjamin-Feir instability in hydrodynamics \cite{Benjamin, Kourakis}.

\section{Localized envelope structures}

Various exact solutions to Eq. (\ref{NLS}) are known \cite{Sulem}, including envelope solitons   \cite{Hasegawa, Dauxois} and breather-type structures \cite{Dysthe}. Interestingly, these have been employed recently in modeling freak-waves (rogue waves).

Envelope solitons, of particular interest to us here, fall into two broad classes: \emph{``bright-type"} and \emph{``dark-type"} solitons \cite{Kourakis, Dauxois} . Bright solutions take the form of a localized region of high intensity and correspond to the case when $Q/P > 0$ -- that is, they can exhibit instability for $\kappa$ within the interval described above. Grey or dark solutions are localized reductions of intensity within a constant ambient background amplitude.  They can arise when $PQ<0$ and are therefore stable under the periodic disturbance described above.

In order to avoid iterative work, we do not provide detailed information on envelope structures, as this can be found elsewhere. Envelope structures are described in full detail in Refs. \onlinecite{Hasegawa} and \onlinecite{Fedele}, and summarized in Ref. \onlinecite{Kourakis}.
Breather-type solutions as models for rogue waves were described e.g. in Ref. \onlinecite{Dysthe}; also see Refs. \onlinecite{Veldes} and \onlinecite{PPCF} for a recent review.

We shall here limit ourselves to pointing out the basic amount of information needed to follow the parametric investigation provided in the following paragraph. In particular, we emphasize that a quantity of crucial importance is the ratio $Q/P$. As shown above, the sign of $Q/P$ determines the  stability profile of modulated wavepackets from a qualitative point of view: a positive (negative) sign implies modulational instability (stability). The value of  $Q/P$ is proportional to the (square) wavenumber $\kappa$; in other words, a perturbation may become unstable in the window $[0, \kappa]$. Furthermore, the ratio $Q/P$ is related to the inverse width of a bright pulse of given amplitude $\psi_0$: to see this, recall that a bright soliton solution of the NLSE (\ref{NLS}) in the form $\psi_0 \, {\rm sech}(\frac{\xi - u_e \tau}{L})$  satisfies the relation $\psi_0  L \sim (P/Q)^{1/2}$. Therefore, for given $\psi_0$ (prescribed i.e. by the lump of energy launched in the system), an envelope soliton will be wider (i.e., larger $L$) if $Q/P$ acquires smaller values; and vice versa. The same is true for dark type solitons, viz. $\psi_0  L \sim |P/Q|^{1/2}$. Recalling that the coefficients $Q$ and $P$ are functions of the carrier wavenumber $k$, we see that the geometric characteristics of envelope solitons will vary from one value of $k$ to another.

The parametric variation of the ratio $Q/P$ will be discussed in the following Section.

\section{Parametric analysis}

In this Section, we shall discuss the parametric dependence of the  coefficients $Q$ and $P$ (and of their ratio) on relevant plasma parameters. We recall that our basic model involved the single parameter $\xi_0$, essentially a function of the (equilibrium) density $n_0$. A comment is required therefore on the density treatment throughout the analysis that follows. This is a one-dimensional setting, so we adopt the Wigner-Seitz (WS) density as a reasonable one-dimensional equivalent to a three-dimensional density. The WS density is formulated as the inverse of the diameter of a sphere whose volume is equal to the mean volume per particle, in three dimensions. In our case, the ``diameter'' of this mean sphere corresponds to the mean separation of the particles in one dimension (assuming each particle to lie in the center of its own ``sphere''). The density is expressed as
\( n_{WS}=\left(\frac{\pi n_{3D}}{6}\right)^{1/3}\).
Now, since the scaling involved the (electron) Fermi energy, which in turn depends on the density, we have had to face the fact that our ``yardstick'' in the plots would be density-dependent. Although this (scaling choice) does not affect our analysis qualitatively, some question might arise on quantitative predictions. To account for this inherent ambiguity in the analysis, one may introduce a (fixed) reference density value, leading to fixed scaling units. We have chosen
$n_0=10^{11} \, m^{-1}$ as a reference density, since this corresponds roughly (by the above formula), in order of magnitude, to representative densities encountered in such places as the interior of a dense dwarf star ($n_{3D} \sim 10^{33}-10^{36} \, m^{-3}$) \cite{Fortov2009}. It is in such environments that the density is thought to be high enough for the manifestation of the relativistic effects under investigation here.

\begin{figure}[h!]
\includegraphics[width=2.9in]{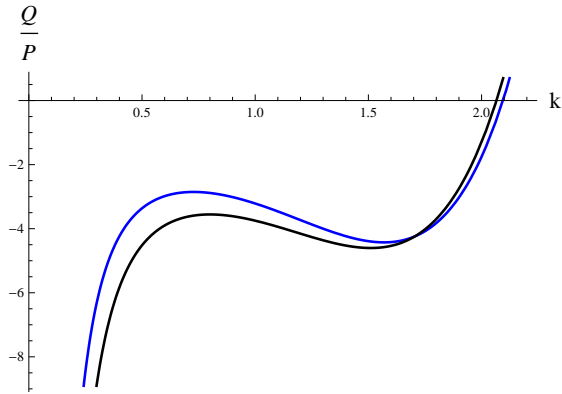}

\caption{(Color online) Plot $Q(k)/P(k)$ for $n_0=10^{11} \, m^{-1}$ \text{(blue)},\,$10^{12} \,m^{-1}$  \text{(black)}.}
\label{figure2}
\end{figure}

Fig. \ref{figure2} depicts $Q/P$ versus the electron equilibrium density. As stated above, we have worked with representative values corresponding roughly to those found in the interior of a white dwarf star.
It is clear that any solution for these two values of the density will be stable for small wavenumbers since the coefficients $P$ and $Q$ are of opposite signs for small $k$- see Equation (\ref{pqsmall}).

\begin{figure}
\includegraphics[width=2.9in]{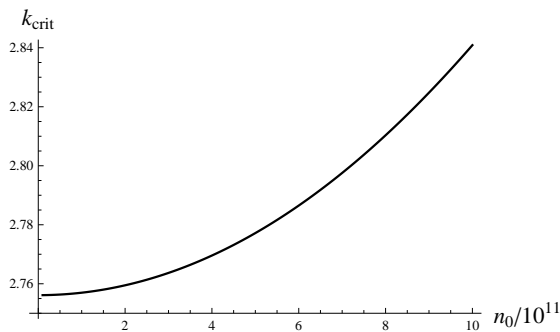}
\caption{(Color online) The carrier wavenumber (instability) threshold, $k_{crit}$, beyond which modulational instability is possible, is shown as a function of the electron equilibrium density, $n_0$ (in \, $m^{-1}$).}
\label{kcrit}
\end{figure}

We have shown that instability sets in when the sign of $P Q$ becomes positive. In practical terms (cf. plots), this occurs above a wavenumber threshold, say $k_{crit}$, corresponding to a root of $Q$ in our case (recall that $P$ is negative here).
Fig. \ref{kcrit} shows the root of $Q$ on the $k$-axis. This root, denoted $k_{crit}$, is the lower bound on the interval over $k$ within which instability can be established. An increase in equilibrium density results in the possibility of instability for lower values of the wavenumber. However, the dependence on density changes (in the interval of interest) is weak. Although the plot shows an increase in $k_{crit}$ with equilibrium density, this is true only because $k_{crit}$ is given in units which depend on density. In real terms one sees a decrease in this critical wavenumber.

\begin{figure}
\includegraphics[width=2.9in]{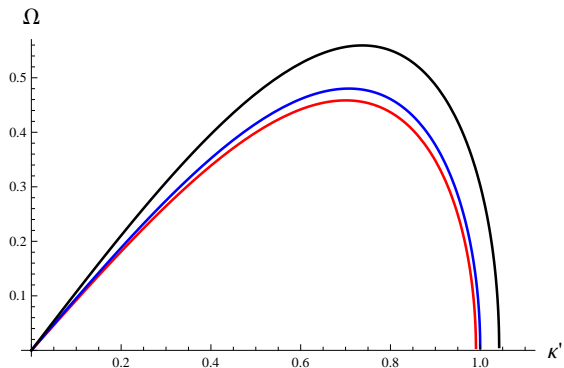}
\caption{(Color online) The growth rate (in plasma frequency units) is shown above as a function of  perturbation wavenumber, $\kappa^\prime=\kappa/\kappa_{crit}|_{n_0=5\times 10^{11}}$. We have taken $k=2.5$ throughout this plot. With reference to the maximum growth rate, the lowest (red), middle (blue) and highest (black) correspond respectively to $n_0=10^{11} \, m^{-1}$, $5\cdot10^{11} \, m^{-1}$ and $10^{12}\, m^{-1}$. }
\label{growthrate}
\end{figure}

The growth rate (\ref{growth}) is depicted in Fig. \ref{growthrate}. For ease of viewing, it has been rescaled so that the middle (blue) curve crosses the $\kappa$-axis at $1$. $\Omega$ is effectively given in units of $\omega_{pi}$ for each curve.  Since $\kappa$ scales as $L_0^{-1}\propto n_0^{-1/2}$, the apparent increase in $\kappa_{crit}$ with equilibrium density is misleading, the true trend (a decrease) being revealed upon restoring dimensions. Therefore the window of instability decreases for higher densities. However, $\Omega$ refers to units of $\omega_{pi}\propto n_0^{1/2}$, so the trend in maximum growth rate (increases with $n_0$) is much stronger in reality. That is to say that there is a larger growth rate over a narrower region of wavenumber for larger equilibrium density.

\section{Conclusions}

We have investigated the modulational dynamics of electrostatic wavepackets in an electron-ion plasma, modeled via a novel relativistic fluid description.
We have shown by adopting a multiscale perturbation methodology that a one-dimensional model of an electron-ion plasma comprising ions and relativistically-degenerate electrons will support envelope structures. The dispersion relation for the carrier wave has been found and the evolution equation for the envelope has been shown to be the nonlinear Schr\"odinger equation.

Our equation of state contains the cold, non-relativistic quantum degeneracy pressure as a low-density limit. Accordingly, our model reduces to the classical (non-relativistic) fluid equations in the appropriate limit (neglecting relativistic effects). 
The only tuneable parameter available is the equilibrium density, since the plasma components were assumed to be ``cold" from the outset, as far as the thermal pressure is concerned; see Equation (1).

Following the paradigm of the Benjamin-Feir instability \cite{Dauxois}, we have found conditions for modulational instability of harmonic-amplitude (Stokes' wave like) solutions. The dependence of these conditions on the wavenumber of the carrier and on equilibrium density have been investigated. We have shown  that the window for instability narrows for higher equilibrium densities, but the maximum growth rate increases.

Our work may be of relevance in white dwarf stars \cite{Koester1990} where the existence of acoustic-type modes has been proposed \cite{BengtShukla,Silvotti},  in which ions would provide the inertia and mainly the electron degeneracy pressure provides the restoring force. Such modes have been predicted \cite{Ostriker}, but haven't been observed to date \cite{Silvotti}. The lack of observations does not imply the absence of acoustic-modes, but may be associated with plasma motion below the detection limit \cite{BengtShukla}. The possibility of the formation of finite amplitude acoustic waves is also suggested in the case of extreme events such as supernova explosions \cite{Fortov2009,BengtShukla}. Various relevant theoretical investigations have been proposed, predicting excitations which are yet to be detected  \cite{Marklund2007,Taibany2012, McKerr, Behery}.

Considering the possibility for experimental confirmation (realization) of our predictions in the laboratory, we note that the present model becomes relevant for ultra-high densities, when both degeneracy and relativistic effects come into play. For instance, an one-dimensional density $n_0 \approx 10^{11} m^{-1}$, corresponding to $\xi_0 \approx 0.1$ (and a 3D equivalent of $n_{3D} \approx 10^{33} m^{-3}$), is not at all inconceivable in view of the already available laser-plasma compression technology \cite{Azechi, Kodama}. Such ultra-dense fermion systems tend to be more ideal (collisionless) in view of the Pauli blocking of electron-electron collisions \cite{Ashcroft}, so that propagating nonlinear structures can be expected to occur.

It might be appropriate, in closing, to discuss the limitations of our work. We have based our analysis on an electrostatic fluid plasma model, which relies on the one-dimensional (1D) equation of state (EoS) (2) above, in order to close the system of fluid equations (3). If one were to assume a 3D geometry, the EoS (2) should be replaced by an appropriate function, as discussed in Refs. \onlinecite{Katsouleas}, \onlinecite{Chavanis} and \onlinecite{MMKPRE2014}. Applying the same perturbation theory would lead to a 3D version of the amplitude equation (\ref{NLS}), which is non-integrable \cite{Sulem, Yang}. As a consequence, if envelope soliton solutions did exist, they would presumably be unstable, and of little value e.g. in real experiments. Furthermore, one might consider going beyond the electrostatic approximation by adding an electromagnetic field. This would be a tedious algebraic task, involving taking into account the full Maxwell's equations. An example of such use of reductive perturbation theory in classical plasmas can be found in earlier work \cite{Veldes, Borhanian1, Borhanian2}. The above lines of research go well beyond our scope in this work, and will not be pursued at this stage.

\acknowledgements
The authors acknowledge support from the EU-FP7 IRSES Programme (grant 612506 QUANTUM PLASMAS FP7-PEOPLE-2013-IRSES). FH and IK gratefully acknowledge support from the Brazilian research fund CNPq (Conselho Nacional de Desenvolvimento Cient\'ifico e Tecnol\'ogico-Brasil).

\begin{appendix}

\section{Analytical expressions for harmonic amplitudes and NLSE (\ref{NLS}) coefficients}

The zeroth harmonic at second order read:
\ba\ptz&=&\frac{-1}{c_1v_g^2-1}\Bigg(v_g\left(\alpha v_g\frac{k^2}{\omega^2}-\frac{2k^3}{\omega^3}\right)-\frac{k^2}{\omega^2}+\nn\\
&&\qquad v_g^2\Bigg[\alpha c_1^2v_g\frac{\omega}{k}-1-\nn\\
&&\alpha \left(\frac{k^2}{\omega^2}-\frac{c_1^2\omega^2}{k^2}\right)\Bigg]\Bigg)\pss=\cad\pss\nn\\
\vtz&=&\left(\frac{1}{v_g}\cad+\frac{k^2}{\omega^2v_g}\right)\pss=\cac\pss\nn\\
\ntz&=&\left(\frac{\cac}{v_g}+\frac{2k^3}{\omega^3v_g}-\alpha\frac{k^2}{\omega^2}\right)\pss=\cab\pss\nn\\
\mtz&=&\left(c_1\cad+\alpha v_gc_1^2\frac{\omega}{k}-1\right)\pss=\cam\pss\nn\\
\utz&=&\left(v_g\cam-\frac{2c_1^2\omega}{k}+\alpha v_gc_1^2\frac{\omega^2}{k^2}\right)\pss=\cau\pss\nn\\
\ay

Second harmonics at second order:
\ba \phi_2^{(2)}&=&-\frac{1}{3k^2\omega^2}\Bigg(\omega^2\Bigg[\frac{1}{2}\left(\alpha c_1^2\frac{\omega^2}{k^2}-1\right)-\nn\\
&&\frac{\alpha}{2}\left(\frac{k^2}{\omega^2}-c_1^2\frac{\omega^2}{k^2}\right)\Bigg]-\frac{3k^4}{2\omega^2}+\frac{\alpha k^2}{2}\Bigg)\psi^2\nn\\
&=&\ccd\psi^2\nn\\
v_{i2}^{(2)}&=&\left(\vfrac\ccd+\frac{k^3}{2\omega^3}\right)\psi^2=\ccc\psi^2\nn\\
n_{i2}^{(2)}&=&\left(\frac{k}{\omega}\ccc+\frac{k^2}{\omega^2}\left(\frac{k^2}{\omega^2}-\frac{\alpha}{2}\right)\right)\psi^2=\ccb\psi^2\nn\\
n_{e2}^{(2)}&=&\left(c_1\ccd+\frac{1}{2}\left(\alpha c_1^2\frac{\omega^2}{k^2}-1\right)\right)\psi^2\nn\\
&=&\ccm\psi^2\nn\\
v_{e2}^{(2)}&=&\frac{\omega}{k}\left(\ccm+c_1^2\left(\frac{\alpha}{2}\frac{\omega^2}{k^2}-1\right)\right)=\ccu\psi^2\nn\\
\ay

The nonlinearity coefficient in Eq. (\ref{NLS}) reads:
\be Q=\frac{\omega}{2(c_1+k^2)}D_3+\frac{k}{2\omega(c_1+k^2)}D_2+\frac{1}{2(c_1+k^2)}D_1 \, \e
where
\ba D_1&=&-\frac{k^2}{\omega}\left(\frac{k}{\omega}\left(\cac+\ccc\right)+\cab+\ccb\right)\nn\\
&&\qquad+\alpha k\left(\cac+\ccc\right)\nn\\
D_2&=&\frac{k^2}{\omega}\left(\frac{3k\alpha}{2\omega}-\cac-\ccc\right)\nn\\
D_3&=&2\xi_0^2\ccd+\frac{\xi_0^2}{c_1}\left(\cam-\ccm\right)+\frac{\xi_0^2}{2c_1}-\nn\\
&& c_1\left(\cam+\ccm\right)+\alpha\Big(\frac{c_1^3\omega^2}{2k^2}+\frac{2c_1\omega^2}{k^2}\ccm+\nn\\
&&\frac{c_1\omega}{k}\left(\cau-\ccu\right)\Big)-\frac{3\alpha}{2}\left(\frac{k^4}{\omega^4}-\frac{c_1^3\omega^2}{k^2}\right)-\nn\\
&&\alpha \left(\frac{k}{\omega}\left(\cac+\ccc\right)-\frac{c_1\omega}{k}\left(\cau+\ccu\right)\right) \, .\ay

\section{Long-wavelength approximation}

We derive approximate expressions for the coefficients $P$ and $Q$ by first approximating their constituents.

\ba \frac{\omega}{k}&=&v_{ph}=\sqrt{\frac{1}{c_1+k^2}}=\sqrt\frac{1}{c_1}\sqrt\frac{1}{1+k^2/c_1}\nn\\
&\approx& \sqrt\frac{1}{c_1}\left(1-\frac{k^2}{2c_1}\right)\nn\\
\omega&\approx&\sqrt\frac{1}{c_1}\left(k-\frac{k^3}{2c_1}\right)\nn\\
v_g&\approx& \sqrt\frac{1}{c_1}\left(1-\frac{3k^2}{2c_1}\right)
\ay

\ba\cad&\approx&\frac{1}{3 k^2}\left(\alpha c_1-3\xi_0^2-4 \right)\nn\\
&\approx&\frac{q_3}{k^2}\nn\\
\cac&\approx&\sqrt{c_1}\frac{q_3}{ k^2}\nn\\
\cab&\approx&c_1\frac{q_3}{k^2}\nn\\
\cam&\approx&\frac{c_1q_3}{k^2}\nn\\
\cau&\approx&\sqrt{c_1}\frac{q_3}{k^2}\ay

\ba\ccd&\approx&\frac{1}{6 k^2}\left(4+3\xi_0^2-\alpha c_1 \right)\nn\\
&\approx&-\frac{q_3}{2k^2}\nn\\
\ccc&\approx&-\sqrt{c_1}\frac{q_3}{2k^2}\nn\\
\ccb&\approx&-c_1\frac{q_3}{2k^2}\nn\\
\ccm&\approx&-c_1\frac{q_3}{2k^2}\nn\\
\ccu&\approx&-\sqrt{c_1}\frac{q_3}{2k^2}\ay

\ba D_1&\approx&\left(\frac{\alpha }{2}-c_1\right)\sqrt{c_1}\frac{q_3}{k}\nn\\
D_2&\approx&-c_1\frac{q_3}{2k}\nn\\
D_3&\approx&\left(\alpha c_1-1\right)\frac{q_3}{2k^2} \, .\ay

Combining the above, we arrive at
\ba Q&\approx&\frac{1}{12c_1^{3/2}k}\left(4+3\xi_0^2-2\alpha c_1\right)\left(4+3\xi_0^2-\alpha c_1\right)\nn\\
P&=&\frac{1}{2}\frac{d^2\omega}{dk^2}\approx-\frac{3k}{2c_1}\sqrt\frac{1}{c_1} \, . \ay

\end{appendix}


\begin{thebibliography}{5}

\bibitem{Giamarchi}  T. Giamarchi, \textit{Quantum  Physics  in  One  Dimension} (Oxford University Press, New York, 2004).

\bibitem{Canuto1} V. Canuto and J. Ventura, \textit{Astrophys. Space. Sci.} {\textbf 18}, 104 (1972)

\bibitem{Canuto2} V. Canuto and J. Ventura, in \textit{Fundamental Cosmic Physics}, edited by C. Gordon and W. Canuto, Vol. 2 (Gordon and Breash Science Publishers, London, 1977), pp.203-353.

\bibitem{Carbonaro} P. Carbonaro, \textit{Nuovo Cimento} {\textbf 103}, 485 (1989).

\bibitem {Koester1990} D. Koester and G. Chanmugam,
\textit{Rep. Prog. Phys.} \textbf{53}, 837
(1990).

\bibitem {Shapiro1983} S. I. Shapiro and S. A. Teukolsky, \textit{Black Holes, White Dwarfs, and Neutron Stars: The Physics of Compact Objects} (Wiley, New York, 1983).

\bibitem {Chabrier2002} G. Chabrier, F. Douchin and A. Y. Potekhin,
\textit{J. Phys. A} \textbf{14}, 9133
(2002).

\bibitem {Fortov2009} V. E. Fortov,
\textit{Phys. Usp.}  \textbf{52},  615
(2009).

\bibitem {Shukla2011} P. K. Shukla and B. Eliasson,
\textit{Rev. Mod. Phys.} \textbf{83}, 885
(2011).

\bibitem{Passoni} M. Passoni, L. Bertagna and A. Zani A., \textit{New J. Phys.} {\bf 12}, 0450122 (2010).

\bibitem{Thiele} R. Thiele, P. Sperling, M. Chen \emph{et al}, \textit{Phys. Rev. E}
\textbf{82}, 056404 (2010).

\bibitem{Shukla} P. K. Shukla and B. Eliasson, \textit{Phys. Rev. Lett.} {\textbf 100}, 036801 (2008).

\bibitem{Barak} G. Barak, H. Steinberg, L. N. Pfeiffer, K. W. West, L. Glazman, F. von Oppen and A. Yacoby,
\textit{Nature Physics} {\textbf 6}, 489 (2010).

\bibitem{Imambekov} A. Imambekov and L. I. Glazman, \textit{Science} {\textbf 323}, 228 (2009).

\bibitem{Haas1} F. Haas, G. Manfredi, P. K. Shukla and P.-A. Hervieux, \textit{Phys. Rev. B} {\textbf 80}, 073301 (2009).

\bibitem{Ghosh} S. Ghosh, N. Chakrabarti and F. Haas, \textit{Europhys. Lett.} {\textbf 105}, 30006 (2014).

\bibitem{McKerr} M. McKerr, I. Kourakis and F. Haas, \textit{Plasma Phys. Controll. Fusion} {\textbf 56}, 035007 (2014).

\bibitem{Tercas} H. Ter\c{c}as, J. T. Mendon\c{c}a and P. K. Shukla, \textit{Phys. Plasmas} {\textbf 15}, 072109 (2008).

\bibitem{Ali1} S. Ali, H. Ter\c{c}as and J. T. Mendon\c{c}a, \textit{Phys. Rev. B} {\textbf 83}, 153401 (2011).

\bibitem{Melrose} D.B. Melrose, M.E. Gedalin, M.P. Kennet and C.S. Fletcher, \textit{J. Plasma. Phys.} \textbf{62}, 233 (1999).

\bibitem{Salpeter}  E. E. Salpeter, \textit{Astrophys. J.} \textbf{134}, 669 (1961).

\bibitem{Eliasson} B. Eliasson and P.K. Shukla, \textit{EPL} \textbf{97}, 15001 (2011).

\bibitem{Akbari} M. Akbari-Moghanjoughi, \textit{Phys. Plasmas} \textbf{17}, 082315 (2010).

\bibitem{Rahman} Ata-Ur-Rahman, S. Ali and A. Mushtaq, \textit{J. Plasma. Phys} \textbf{79}, 817 (2013).

\bibitem{MMKPRE2014} M. Mc Kerr, F. Haas, I. Kourakis, \textit{Physical Review E} \textbf{90}, 033112  (2014).

\bibitem{HaasPPCF}  F. Haas and I. Kourakis, \textit{Plasma Phys. Cont. Fusion} \textbf{57}, 044006 (2015).

\bibitem{Chandrasekhar} S. Chandrasekhar, \textit{Mon. Not. R. Astron. Soc.} {\bf 95}, 207 (1935).

\bibitem{Katsouleas} T. Katsouleas and W.B. Mori, \textit{Phys. Rev. Lett.} \textbf{61}, 90 (1988).

\bibitem{Haasbook} F. Haas, \textit{Quantum Plasmas: an Hydrodynamic Approach} (Springer, New York, 2010).

\bibitem{Chavanis} P.-H. Chavanis, \textit{Phys. Rev. D} \textbf{76}, 023004 (2007).

\bibitem{Kourakis} I. Kourakis and P.K. Shukla, \textit{Nonlin. Proc. Geophys.} \textbf{12}, 407 (2005).

\bibitem{Benjamin} T. B. Benjamin and J.E. Feir, \textit{J. Fluid. Mech} \textbf{27}, 417 (1967).

\bibitem{Sulem} C. Sulem and P-L. Sulem, \textit{The Nonlinear Schr\"odinger Equation} (Springer, 1993).

\bibitem{Hasegawa} A. Hasegawa, \textit{Optical Solitons in Fibers} (Springer-Verlag, 1989).

\bibitem{Dauxois} T. Dauxois and M. Peyrard, \textit{Physics of Solitons} (Cambridge University Press, 2006).

\bibitem{Dysthe} K. Dysthe and K. Trulsen, \textit{Phys. Scr.} \textbf{T82}, 48 (1999).

\bibitem{Fedele} R. Fedele and H. Schamel, \textit{Eur. Phys. J. B} \textbf{27}, 313 (2002).

\bibitem{Fedele-Schamel} R. Fedele, H. Schamel and P.K. Shukla, \textit{Phys. Scripta} \textbf{T98}, 18 (2002).

\bibitem{Veldes} G.P. Veldes, J. Borhanian, M. McKerr, V. Saxena, D.J. Frantzeskakis and I. Kourakis, \textit{J. Opt}
\textbf{15}, 064003 (2013).

\bibitem{PPCF} M. McKerr, I. Kourakis and F. Haas, \textit{Plasma Phys. Controll. Fusion} \textbf{56}, 35007 (2014).

\bibitem{BengtShukla}  B. Eliasson and P. K. Shukla,
\textit{Europhys. Lett.} \textbf{97},  15001  (2012).

\bibitem{Silvotti} R. Silvotti, G. Fontaine, M. Pavlov, M. Pavlov, T. R. Marsh, V. S. Dhillon, S. P. Littlefair and F. Getman,
\textit{Astron. Astrophys.} \textbf{525}, A64 (2011).

\bibitem{Ostriker} J. P. Ostriker,
\textit{Ann. Rev. Astron. Astrophys.} \textbf{9}, 353 (1971).


\bibitem{Marklund2007} M. Marklund, B. Eliasson and P. K. Shukla,
\textit{Phys. Rev. E}  \textbf{76}, 067401 (2007).

\bibitem {Taibany2012} W. F. El-Taibany and A. A. Mamun,
\textit{Phys. Rev. E} \textbf{85},  026406 (2012).

\bibitem{Behery} E. E. Behery, F. Haas and I. Kourakis, \textit{Phys. Rev. E} \textbf{93}, 023206 (2016).

\bibitem{Azechi} H. Azechi, T. Jitsuno, T. Kanabe, M. Katayama, K. Mima, N. Miyanaga, M. Nakai, S. Nakai, H. Nakaishi, M. Nakatsuka, A. Nishiguchi, P. A. Norrays, Y. Setsuhara, M. Takagi, M. Yamanaka and C. Yamanaka,  \textit{Laser Part. Beams}, \textbf{9}, 193 (1991).

\bibitem{Kodama}  R. Kodama, P. A. Norreys, K. Mima, A. E. Dangor, R. G. Evans, H. Fujita, Y. Kitagawa, K. Krushelnick, T. Miyakoshi, N. Miyanaga, T. Norimatsu, S. J. Rose, T. Shozaki, K. Shigemori, A. Sunahara, M. Tampo, K. A. Tanaka, Y. Toyama, T. Yamanaka and M. Zepf,
\textit{Nature}, \textbf{412}, 798 (2001).

\bibitem{Ashcroft} N. W. Ashcroft and N. D. Mermin, \textit{Solid State Physics}  (Orlando, Saunders College Publishing, 1976).

\bibitem{Yang} J. Yang, \emph{Nonlinear Waves in Integrable and Nonintegrable Systems} (SIAM, USA).

\bibitem{Borhanian1}  J. Borhanian, I. Kourakis and S. Sobhanian,
Phys. Lett. A \textbf{373}, 3667  (2009). 

\bibitem{Borhanian2}  J. Borhanian, S. Sobhanian, I. Kourakis and A. Esfandyari-Kalejahi,
Phys. Plasmas {\bf 15}, 093108   (2008). 

\end{thebibliography}
\end{document}